\documentclass[preprint,aps,prd,amsmath,amssymb]{revtex4}
\usepackage{graphicx}
\usepackage{color}
\usepackage[latin1]{inputenc}
\usepackage{ulem}
\newcommand{\be}{\begin{eqnarray}}
\newcommand{\beq}{\begin{equation}}
\newcommand{\eeq}{\end{equation}}
\newcommand{\ee}{\end{eqnarray}}

\newcommand{\bmp}{\noindent\begin{minipage}{16cm}}
\newcommand{\emp}{\end{minipage}\vskip 7mm} 

\usepackage{graphicx}
\usepackage{dcolumn}
\usepackage{bm}
\usepackage{amsmath}
\usepackage{amsfonts}
\usepackage{bbm}
\usepackage{subfigure}

\def\drawbox#1#2{\hrule height#2pt
        \hbox{\vrule width#2pt height#1pt \kern#1pt
              \vrule width#2pt}
              \hrule height#2pt}

\def\Asym#1#2{\vcenter{\vbox{\drawbox{#1}{#2}
              \kern-#2pt 
              \drawbox{#1}{#2}}}}

\begin{document}
\title{Dark Matter from new Technicolor Theories}
\author{Sven Bjarke {\sc Gudnason}}
\email{gudnason@nbi.dk}
\author{Chris {\sc Kouvaris}}
\email{kouvaris@nbi.dk}
\author{Francesco {\sc Sannino}}
\email{sannino@nbi.dk} \affiliation{The Niels Bohr Institute,
Blegdamsvej 17, DK-2100 Copenhagen \O, Denmark }


\begin{abstract}
We investigate dark matter candidates emerging in recently proposed technicolor theories. We determine the relic density
of the lightest, neutral, stable technibaryon having imposed weak
thermal equilibrium conditions and overall electric neutrality of the Universe. In addition we consider
sphaleron processes that violate baryon, lepton and technibaryon number. Our analysis is performed in the case
of a first order electroweak phase transition as well as a second order one. We argue that, in both cases, the
new technibaryon contributes to the
  dark matter in the Universe. Finally we examine the problem of the constraints on these types of
  dark matter components from earth based experiments.

\end{abstract}


\maketitle

\section{Introduction}

The origin of the dark matter is one of the most intriguing open problems of modern particle physics and
cosmology. A few decades ago, astronomers realized that it was impossible to explain the motion of galaxies,
just by accounting the luminous part of the galaxy. The most important observational evidence comes from the
rotation curves of spiral galaxies. Astronomers are able to estimate the velocities of clouds of hydrogen atoms,
just by looking at the Doppler shifted 21 cm emission. Naively one would expect by using Newton's law, that the
velocity of clouds like these should fall as $\propto r^{-1/2}$, when $r$ becomes larger than the radius of the
luminous part of the galaxy ($r$ being the distance to the center of the galaxy). However, astronomers observe
an almost constant velocity, independent of $r$. Unless one proposes a radical change of the laws of gravity or
of motion, it is logical to assume that matter that does not interact ``too much'' and therefore appears dark to
us, has to fill the space of the galaxy.

 Once we assume that a type of dark matter (DM) is
responsible for the discrepancies of the motion of the galaxies, then there are two distinct
possibilities for the origin of the DM. The first type of candidates are of baryonic origin and they are called
MACHO (Massive Compact Halo Objects). The MACHO can be brown dwarf stars, giant planets and massive black holes.
Massive black holes with masses near $100M_{\bigodot}$ can be remnants at the center of supernovae explosions.
The brown dwarf stars are stellar objects with masses less than $0.1M_{\bigodot}$. Since a proton star needs at
least a mass of $0.1M_{\bigodot}$ to ignite nuclear fusion, brown dwarfs never get to begin the nuclear fusion
of hydrogen. Giant planets can be also a component of MACHOs with masses of the order of $0.001M_{\bigodot}$.
However, observations so far showed with a high level of confidence that MACHOs cannot account for more than
$20\%$ of the DM \cite{Alcock:2000ph}.

The second possibility is to have matter of nonbaryonic origin. In contrast with MACHOs that are compact objects
of baryonic matter, in this case we have particles that are neutral and interact only through gravitational and
weak forces. The name WIMP (Weakly Interacting Massive Particles)  is frequently used for these particles. The
Standard Model does not have the particles with the desired properties. Ordinary neutrinos are the only
electrically neutral objects that interact weakly within the SM. They are, however, too light and would compose
part of a {\it hot}-type dark matter, which is usually not considered in a viable cosmological model. This is so
since hot dark matter would smear out large scale structure of galaxies. Supersymmetry, for example, provides
some natural WIMP candidates for cold dark matter such as neutralinos.

In the search of WIMP candidates, particles related to technicolor
theories were also investigated as possible sources of cold DM. This
idea was pioneered by Nussinov in
\cite{Nussinov:1985xr,Nussinov:1992he} and further investigated in
\cite{Barr:1990ca}. He imagined that DM could be accounted by a
technibaryon asymmetry which can be ultimately related to the
ordinary baryon asymmetry. Although it was clear that technicolor
can give ``reliable'' DM candidates, little interest was shown in
the past because of the severe problems that most technicolor
theories suffer from such as large flavor changing neutral current
processes and/or problems with the Electroweak Precision
Measurements. Progress in this direction has been made recently. We
have constructed explicit technicolor extensions of the SM passing
the precision tests \cite{Sannino:2004qp,Hong:2004td,
Dietrich:2005jn,Gudnason:2006ug}. Because of their walking nature
they have very much reduced flavor changing neutral current
processes. Therefore it does make sense to revisit the possibility
of technibaryons as components of DM. We have started this analysis
in \cite{Gudnason:2006ug} for the models with technifermions in the
adjoint representation of the technicolor gauge group. In this case
the technibaryon is one of the would-be Goldstone bosons of the
underlying technicolor theory. In Ref.~\cite{Appelquist:2003hn} has
been also suggested that technicolor theories may lead to dark
matter candidates of similar nature.

The goal of the present work is threefold: First, we will provide the basic set up while showing the detailed
computations needed to determine the present relic density of our technibaryon DM candidate. Second, we
consider two different orders of the electroweak phase transition and compare the results. Third, using our
low energy effective theory developed in \cite{Gudnason:2006ug} we compute the relevant cross sections needed to
examine the problem of the constraints on these types of DM components from earth based experiments.

In the next section we review the basic properties of the new technicolor theory and present the lightest
neutral technibaryon (LTB) of the theory. In section III we present our main calculation of the technibaryon
contribution to DM. In section IV we investigate the
experimental constraints and comment on the detection of our technibaryon WIMP. Finally we briefly conclude in section V.

\section{Conventions and notations.}
The minimal walking technicolor model
\cite{Sannino:2004qp,Hong:2004td, Dietrich:2005jn,Gudnason:2006ug}
has two techniflavors (techni-up $U$, and techni-down $D$)
transforming according to the adjoint representation of the $SU(2)$
technicolor gauge group. The global flavor symmetry is $SU(4)$ which
breaks spontaneously to $SO(4)$. The associated low energy effective
theories have been constructed in \cite{Gudnason:2006ug}. There are
9 Goldstone bosons, 3 of which are eaten by the $W$ and $Z$ bosons
and are the technicolor equivalent of the ordinary pions. Three of
the remnant six Goldstone bosons
 transform under techniflavor symmetries as:
 \beq
\begin{array}{ccc}
U_LU_L \ , & \quad  D_LD_L  \ , & \quad  U_LD_L \ ,
 \label{othergbs}
\end{array}
 \eeq
  with electric charges, respectively
 \beq y +1, \qquad y-1, \qquad y \ ,
 \eeq
 while the other three are the antiparticles. In the following we  drop the subscript $L$ when referring to the above states.
The parameter $y$ can take any real value. It is related to the hypercharge of the techniquarks and is such that
gauge anomalies cancel out. Additionally in order to cancel Witten's global anomaly we simply add an extra
family of leptons
 \beq
 {\cal L}_L = \left( \begin{array}{c} \nu' \\ \zeta
\end{array} \right)_L , \qquad \left(\nu'_R \ ,~\zeta_R \right) \ ,
\label{newlepton}
 \eeq
 with hypercharges:
\beq -\frac{3y}{2} \ , \qquad \left(\frac{-3y+1}{2}
 \ , \frac{-3y-1}{2} \right) \ , \eeq
where we use the convention:
\beq Q = T_3 + Y \ . \eeq

{}A typical cold DM candidate must be electrically neutral and at most have weak interactions. {}For example we
can choose $y=1/3$, i.e. the SM-like hypercharge assignment and in this case  $\nu'$ (the \emph{new neutrino})
is electrically neutral and can be made stable by requiring no mixing
 with the lighter neutrino species. This case is similar to the one
 studied in \cite{Rybka:2005vv}.
In this paper we consider the case with $y=1$, where the second
(technibaryon) Goldstone boson of Eq.~(\ref{othergbs}) is now electrically neutral. One the other hand, the
\emph{new leptons} $\nu'$ and $\zeta$ have electric charges $-1$ and $-2$, respectively. The
still not directly observed Goldstone bosons will acquire
masses via new interactions. If we
assume these
interactions to preserve the technibaryon number, then the
electrically neutral $DD$, if it is the lightest
technibaryon, is stable. We will denote with LTB the lightest
 neutral technibaryon particle. We conclude that
$DD$ is an interesting candidate which can be a cold DM component.

\section{Computing the LTB relic density.}

We now explicitly compute the $DD$-type boson relic density in the case it is neutral and stable. We impose
thermal equilibrium and overall electric neutrality for the matter in the Universe.
Imposing overall electrical neutrality avoids the huge energetic Coulomb costs due to electric fields of the
otherwise uncanceled charges in the Universe. In addition to the theoretical reasons, observations confirm an
overall neutrality. Thermal equilibrium occurs among different particles as long their rate $\Gamma$ of
interaction is much larger than the the expansion rate of the Universe $H$, where $H$ is the Hubble constant. If
$H>\Gamma$ at a given time the particles decouple from each other and hence can no longer be in thermal
equilibrium.

At some energy scale higher than the electroweak one, following the work of Nussinov \cite{Nussinov:1985xr}, we assume the existence of a mechanism leading to a technibaryon asymmetry in the Universe. Given that the
technibaryon and baryon number have a very similar nature such an asymmetry is very plausible and can have a common
origin. Here we will not speculate further on the origin of the (techni)baryon asymmetry but will relate it to
the observed baryon asymmetry as done by Nussinov as well as in \cite{Barr:1990ca}. Here we provide detailed
computations for our specific technicolor model for two types of electroweak phase transitions.

Even if one is able to produce an asymmetry above or around the electroweak scale the (techni)baryon number is
spoiled by quantum anomalies. Fortunately although the baryon ($B$),
technibaryon ($TB$), lepton ($L$) number and the new lepton number for
the new lepton family ($L^{\prime}$) are not conserved
individually, their differences, i.e. $B-L$ and
$3TB-L$ are
preserved. This fact allows for a nonzero
(techni)baryonic asymmetry to survive. The processes leading to such a
violation are termed ``sphaleron''
processes and at the present time are negligible. However these
processes were active during the time the
Universe had a temperature above or at the scale of the electroweak
symmetry breaking ($\sim$ $250$ GeV). Indeed
these processes were rapid enough to thermalize baryons, leptons and
technibaryons. At some point as the
Universe expands and its temperature falls, the
baryon-lepton-technibaryon violation processes cease to be
significant. The precise value of this temperature $T^{\ast}$  depends
on the underlying theory driving
electroweak symmetry breaking. Within the SM framework and assuming
the validity of the semiclassical
calculation of the tunneling effect \cite{Kuzmin:1985mm}, $T^*$ has
been estimated by equating the rate of the
sphaleron processes to $H$. According to \cite{Kuzmin:1985mm}, $T^*$
satisfies the following equation \beq
 T^* = \frac{2M_W(T^*)}{\alpha_W
 \ln\left(\frac{M_{Pl}}{T^*}\right)}B\left(\frac{\lambda}{\alpha_W}\right) \ ,
\label{tunnel}
 \eeq
 where $M_W$ is the mass of the $W$ bosons, $M_{Pl}$ is the
Planck scale, $\alpha_W$ is the weak coupling constant, $\lambda$ is
 the self coupling of the Higgs boson and
$B(\lambda/\alpha_W)$ is a function that takes on values from 1.5 to 2.7 as the ratio $\lambda/\alpha_W$ goes
from zero to infinity \cite{Kuzmin:1985mm, Klinkhamer:1984di}. As we already mentioned this formula is an
approximation and it depends on the not very well known ratio  of $\lambda/\alpha_W$. According to what is the
value of this ratio, $T^*$ can vary within the $150-250$ GeV
 range. In
technicolor theories, since the Higgs is a composite object, the self-coupling $\lambda$ is in principle
calculable. An estimation $\lambda = 1/8$ for our specific model was given in \cite{Dietrich:2005jn}. Since
$\alpha_W = 1/29$ (or a bit larger at the electroweak scale), the ratio $\lambda/\alpha_W$ gives a $T^*$ around
$200$ GeV.

It is time to introduce now the chemical potentials for the relevant particle species. We here follow Ref.
\cite{Harvey:1990qw}
\begin{align}
\mu_W \qquad & {\rm for } \quad W^- & \mu_{dL} \qquad & {\rm for }
\quad d_L, s_L, b_L\nonumber \\
\mu_0 \qquad & {\rm for } \quad \phi^0 & \mu_{dR} \qquad & {\rm
for }
\quad d_R, s_R, b_R \nonumber \\
\mu_- \qquad & {\rm for } \quad \phi^- & \mu_{iL} \qquad & {\rm
for }
\quad e_L, \mu_L, \tau_L \nonumber \\
\mu_{uL} \qquad & {\rm for } \quad u_L, c_L, t_L & \mu_{iR} \qquad
&
   {\rm for } \quad e_R, \mu_R, \tau_R \nonumber \\
\mu_{uR} \qquad & {\rm for } \quad u_R, c_R, t_R & \mu_{\nu iR}
\qquad
& {\rm for } \quad \nu_{eR}, \nu_{\mu R}, \nu_{\tau R} \nonumber \\
\mu_{\nu iL} \qquad & {\rm for } \quad \nu_{eL}, \nu_{\mu L},
\nu_{\tau L} \nonumber
\end{align}
where the indices $L$ and $R$ denote chirality. We have a common chemical potential for the up, charm and top
quarks, and a different one for the other triplet of down, strange and bottom. A common chemical potential has
to do with the fact that at the scale of interest QCD interactions put quarks of the same charge on equal
footing. We introduce a different chemical potential for all
of the leptons. Also in order to be
as general as
possible we have assumed the existence of right handed neutrinos and
introduced different chemical potentials
for the left and the right handed particles. The thermal equilibrium
conditions associated to the weak
interactions read:
\begin{align}
\mu_W &= \mu_- + \mu_0 &\qquad (W^- &\leftrightarrow \phi^- +
\phi^0) \ ,
\label{smproc1}\\
\mu_{dL} &= \mu_{uL} + \mu_W &\qquad (W^- &\leftrightarrow \bar
u_L +
d_L) \ , \label{smproc2}\\
\mu_{iL} &= \mu_{\nu iL} + \mu_W &\qquad (W^- &\leftrightarrow
\bar\nu_{iL} +
e_{iL}) \ , \label{smproc3}\\
\mu_{\nu iR} &= \mu_{\nu iL} + \mu_0 &\qquad (\phi^0
&\leftrightarrow
\bar\nu_{iL} + \nu_{iR}) \ , \label{smproc7}\\
\mu_{uR} &= \mu_0 + \mu_{uL} &\qquad (\phi^0 &\leftrightarrow \bar
u_L +
u_R) \ , \label{smproc4}\\
\mu_{dR} &= -\mu_0 + \mu_W + \mu_{uL} &\qquad (\phi^0
&\leftrightarrow
d_L + \bar d_R) \ , \label{smproc5}\\
\mu_{iR} &= -\mu_0 + \mu_W + \mu_{\nu iL} &\qquad (\phi^0
&\leftrightarrow e_{iL} + \bar e_{iR}) \ , \label{smproc6}
\end{align}
where it is understood that the Higgs is a composite of two techniquarks. The Goldstone bosons of
Eq.~(\ref{othergbs}) are gauged under the weak symmetry and hence we introduce the following chemical potential
for these Goldstone bosons and the new lepton family of Eq.~(\ref{newlepton})
\begin{align}
\mu_{\zeta L} \qquad &{\rm for} \quad \zeta_L & \mu_{UU} \qquad
&{\rm
  for} \quad UU \nonumber \\
\mu_{\zeta R} \qquad &{\rm for} \quad \zeta_R & \mu_{UD} \qquad
&{\rm
  for} \quad UD \nonumber \\
\mu_{\nu'L} \qquad &{\rm for} \quad \nu_{\zeta L} & \mu_{DD}
\qquad
  &{\rm for} \quad DD \nonumber\\
\mu_{\nu'R} \qquad &{\rm for} \quad \nu_{\zeta R} \nonumber
\end{align}
The corresponding thermal equilibrium equations for the extra
particles introduced by the technicolor theory per se are
\begin{align}
\mu_{\zeta L} &= \mu_W + \mu_{\nu'L} &\qquad (\zeta_L
&\leftrightarrow
W^- + \nu_{\zeta L}) \ , \label{tcproc1} \\
\mu_{UD} &= \mu_{DD} - \mu_W &\qquad (DD &\leftrightarrow
UD + W^-) \ , \label{tcproc2} \\
\mu_{UU} &= \mu_{UD} - \mu_W = \mu_{DD} - 2\mu_W &\qquad (UD
&\leftrightarrow UU + W^-)  \ , \label{tcproc3} \\
\mu_{\zeta R} &= -\mu_0 + \mu_{\zeta L} &\qquad (\phi^0
&\leftrightarrow \zeta_L + \bar\zeta_R) \ ,\label{tcproc4} \\
\mu_{\nu'R} &= \mu_0 + \mu_{\nu'L} &\qquad (\phi^0
&\leftrightarrow \bar\nu_{\zeta L} + \nu_{\zeta R}) \ ,
\label{tcproc5}
\end{align}
where Eq.~(\ref{tcproc2}) has been used in Eq.~(\ref{tcproc3}).

Each classical gauge and scalar field configuration with a given
topological number leads to a simultaneous jump for  {\it all} of
the anomalous charges. Hence each quark-doublet generation,
lepton-doublet generation, the new lepton family number as well as
techniquark number are violated by
the same classical field
configuration. The one loop anomalous coefficient dictates the
relative amount of the jump for each anomalous charge when turning
on a given classical field configuration.

With the normalization of $1/3$ for the technibaryonic charge for
our techniquarks and $1/3$ for the ordinary quark-baryonic charge of
the quarks, $1$ for all of the leptons, the simplest classical
configuration with one unit of topological charge will induce a
transition from the the vacuum of the theory to a state containing
one baryon (per each generation), one lepton (for each generation),
a technibaryon-like object with 3 technibaryons and one new lepton.
The relation among the chemical potentials emerging from the above
is:
\begin{eqnarray}
3(\mu_{u_L} +2\mu_{d_L}) +\mu + \frac{1}{2}\mu_{UU} + \mu_{DD}
+\mu_{\nu^{\prime}}=0 \ .
\end{eqnarray}
The parameter $\mu$ is defined as $\sum_i\mu_{\nu iL} \equiv \mu$.
We have assumed that the difference in the baryon number between two
different quark-doublet generations is created identical before the
electroweak phase transition. A similar relation will be assumed for
the lepton charges. Note that the difference is not affected by the
weak anomaly and hence will not be generated later on.

We can now turn to the calculation of number densities. The
difference between the number densities of particles and their
corresponding antiparticles is given by
\begin{align}
n &= n_+ - n_- =
g\int\frac{d^3k}{(2\pi)^3}\frac{1}{z^{-1}e^{E\beta} - \eta} -
g\int\frac{d^3k}{(2\pi)^3}\frac{1}{ze^{E\beta} - \eta} \ ,
\end{align}
where $n_+$ and $n_-$ are the number densities of the particles and antiparticles, respectively. The constant
$g$ is the multiplicity of the degrees of freedom (spin for example), $\beta = 1/T$ in units $k_B = 1$, and
$\eta$ takes on the values $1$ and $-1$ for bosons and fermions respectively. The fugacity $z = e^{\mu \beta}$
and $E$ is the energy. The ratio $\mu/T$ in the Universe is sufficiently small that we can Taylor expand the
above relation. The number density now can be written as \beq n =
\begin{cases} gT^3\frac{\mu}{T}f\left(\frac{m}{T}\right)
& \text{for fermions} \ , \\
gT^3\frac{\mu}{T}g\left(\frac{m}{T}\right) & \text{for bosons} \ ,
\end{cases}
\label{fgfunctions} \eeq
where the functions $f$ and $g$ are defined as follows
\begin{align}
f(z) &= \frac{1}{4\pi^2}\int_0^\infty dx\
x^2\cosh^{-2}\left(\frac{1}{2}\sqrt{x^2+z^2}\right) \ , \label{funcf} \\
g(z) &= \frac{1}{4\pi^2}\int_0^\infty dx\
x^2\sinh^{-2}\left(\frac{1}{2}\sqrt{x^2+z^2}\right) \ . \label{funcg}
\end{align}
We now differentiate two cases according to the order of the electroweak phase transition. In the case of a
second order or weak first order electroweak phase transition we expect that the temperature $T^*$ is below the
temperature of the phase transition. This means that baryon, lepton and technibaryon violating processes persist
after the phase transition. The second possibility is to have a strong first order phase transition where the
violating processes freeze right at the phase transition. We are going to examine separately the two different
cases.

Assuming that the violating processes persist even after the phase transition, we need to impose two conditions
here: Electric neutrality and set $\mu_0=0$, since the Higgs boson condenses and the electroweak symmetry breaks
spontaneously. Recall that we can introduce a nonzero chemical potential only for unbroken symmetries whose
generators commute with all of the gauge ones. Here the Higgs boson is a composite particle, made of techni-up
and techni-down quarks $(\bar{U}U + \bar{D}D)/\sqrt{2}$. Therefore when we refer to $\mu_0$ as the chemical
potential, we mean the chemical potential of the composite object.

From Eq.~(\ref{fgfunctions}) we see that the number densities, in the leading approximation, are linear in the
chemical potential for small $\mu/T$. For convenience we express the baryon number density as
 \beq B \equiv \frac{n_{B} - n_{\bar B}}{gT^2/6} \ . \eeq
 We shall use the same normalization (dividing the number density by $gT^2/6$) also for the lepton number,
technibaryon number etc. Since in the end we care only for ratios of number densities, the normalization
constant cancels out.

We conveniently define the function $\sigma$ as follows
\beq \sigma_i = \begin{cases}
6f\left(\frac{m_i}{T^*}\right) & \textrm{for fermions} \ , \\
6g\left(\frac{m_i}{T^*}\right) & \textrm{for bosons} \ ,
\end{cases} \eeq
where $f$ and $g$ are those of Eq.~(\ref{funcf}) and
(\ref{funcg}), respectively and the index $i$ refers to the particle
in question.

For all of the SM particles the statistical function is taken to be 1 and 2 for massless fermions and bosons
respectively, except for the top quark which we treat massive as $m_t$ is of order $T^*$. The reason why we can
take the other SM particles to be massless in the statistical function is that $m \ll T^*$. However, the
technibaryons as well as the particles of the new lepton family have masses that cannot be ignored. We should
emphasize that we calculate the baryon and lepton numbers at the temperature $T^*$ where sphalerons die out.

The baryon density can be written as
\begin{align}
B &= \frac{3}{3}\left[(2 + \sigma_t)(\mu_{uL} + \mu_{uR}) +
3(\mu_{dL} + \mu_{dR})\right] \ ,
\nonumber \\
&= (10 + 2\sigma_t)\mu_{uL} + 6\mu_W + (\sigma_t - 1)\mu_0 \ ,
\end{align}
where Eqs.~(\ref{smproc2}), (\ref{smproc4}) and (\ref{smproc5}) have been used and the factor in the first line
includes number of colors and that the baryon number of each quark is $1/3$. The factor 3 of the down-type quarks is
the number of families and equivalent the factor $2+\sigma_t$ is the number of families taking into account the
top mass effect.

Similarly the lepton number for the Standard Model leptons is
\begin{align}
L &= \sum_i(\mu_{\nu iL} + \mu_{\nu iR} + \mu_{iL} + \mu_{iR}) \ ,
\nonumber \\
&= 4\mu + 6\mu_W \ .
\end{align}
For the new lepton family we have
\begin{align}
L' &=  \sigma_\zeta(\mu_{\zeta L} + \mu_{\zeta R})
 + \sigma_{\nu'}(\mu_{\nu'L} + \mu_{\nu'R}) \ , \nonumber\\
&= 2(\sigma_{\nu'} + \sigma_\zeta)\mu_{\nu'L}
+ 2\sigma_{\zeta}\mu_W + (\sigma_{\nu'} - \sigma_\zeta)\mu_0 \ .
\end{align}
Similarly for the technibaryons we get
\begin{align}
TB &= \frac{2}{3}(\sigma_{UU}\mu_{UU} + \sigma_{UD}\mu_{UD} +
\sigma_{DD}\mu_{DD}) \ ,
\nonumber \\
&= \frac{2}{3}(\sigma_{UU}+\sigma_{UD}+\sigma_{DD})\mu_{DD} -
\frac{2}{3}(\sigma_{UD} + 2\sigma_{UU})\mu_W \ .
\end{align}
%
The charge constraint for all the particles is
\begin{align}
Q =&\  \frac{2}{3}\cdot3(2 + \sigma_t)(\mu_{uL} + \mu_{uR}) -
\frac{1}{3}\cdot3\cdot3(\mu_{dL} + \mu_{dR}) \nonumber \\
&- \sum_i(\mu_{iL} + \mu_{iR}) - 2\cdot2\mu_W - 2\mu_- \nonumber \\
&+ 2\sigma_{UU}\mu_{UU} + \sigma_{UD}\mu_{UD} - 2\sigma_\zeta(\mu_{\zeta L} +
\mu_{\zeta R}) - \sigma_{\nu'}(\mu_{\nu'L} + \mu_{\nu'R}) \ .
\end{align}
For the
first order phase transition we will need also the neutrality with respect to the weak isospin charge which is
\begin{align}
Q_3 =&\ \frac{3(2+\sigma_t)}{2}\mu_{uL} -
\frac{3\cdot3}{2}\mu_{dL} + \frac{1}{2}\sum_{i=1}^3(\mu_{\nu iL} -
\mu_{iL}) - 4\mu_W - (\mu_0 + \mu_{-}) \nonumber \\
&\ + (\sigma_{UU}\mu_{UU} - \sigma_{DD}\mu_{DD}) +
\frac{1}{2}(\sigma_{\nu'}\mu_{\nu'L} - \sigma_{\zeta}\mu_{\zeta
L}) \ . 
\end{align}
The need for the isospin neutrality condition, in the first order case, comes from the fact that we are
computing our final relic densities above the electroweak phase transition where the weak isospin is unbroken.

Since it is not clear whether the electroweak phase transition is first or second order, we should examine both
cases. It is expected as in \cite{Harvey:1990qw} that a strong first order phase transition occurs fast enough
to ``freeze'' the baryon and technibaryon violating processes just at the transition. In this case one
calculates the equilibrium conditions just before the transition. On the other hand, in a second order phase
transition we expect the violating processes to persist below the phase transition and the equilibrium
conditions are imposed after the phase transition. If the phase diagram as function of temperature and density
of our technicolor theory would be known a specific order of the electroweak phase transition would be used.

When the ratio between the number densities of the technibaryons to the baryons is determined we have \beq
\frac{\Omega_{TB}}{\Omega_B} = \frac{3}{2}\frac{TB}{B}\frac{m_{TB}}{m_p} \ , \eeq here $m_{TB}$ is the mass of the LTB (the
$m_{DD}$) and $m_p$ is the mass of the proton.

Note that a possible mixing between the new family and an ordinary standard model family would dilute the relative $\nu'$ abundance and eventually annihilate $L' \ $.

\subsection{2nd Order Phase Transition}
\begin{figure}[!tbp]
\begin{center}
\includegraphics[width=0.7\linewidth]{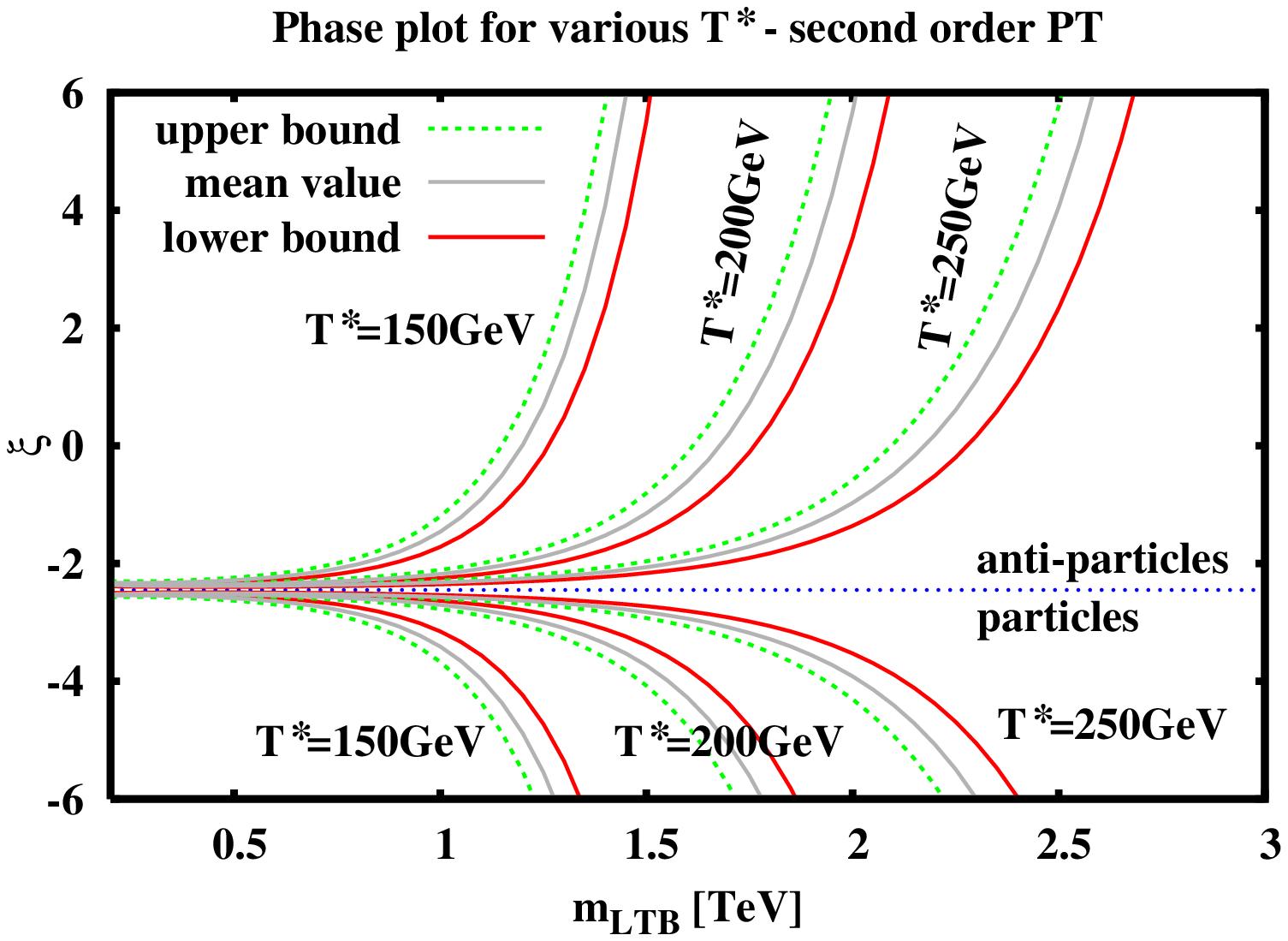}
\caption{Plot representing the region of the parameters according to which the fraction of technibaryon matter
    density over the baryonic one takes on the values $[3.23,\,5.55]$.
We consider a second order phase transition.
    The parameters in the plot are the mass of the LTB DM particle and
    $\xi$ of Eq.~(\ref{soxidef}). The plot includes various values of
    $T^{\ast}$. The dotted line separates
    areas of abundant particles
    and anti-particles. }
\label{fig:Omega}
\end{center}
\end{figure}

Here the two conditions we have to impose are: Overall electrical
neutrality and $\mu_0=0$ for the chemical
potential of the Higgs boson.
The ratio between the number density of the technibaryons to the
baryons can be expressed as function of the $L/B$ and
$L^{\prime}/B$ ratios. In order to provide a simple
and compact
expression we consider the limiting case in which the $UU$ and
$UD$ technibaryons are substantially heavier than the $DD$
companion, the top is light with respect to the electroweak phase
transition temperature and the new lepton family is degenerate,
i.e. $\sigma_{\zeta}=\sigma_{\nu^{\prime}}$. In this approximation
the ratio simplifies to
\begin{eqnarray}
-\frac{TB}{B}=\frac{\sigma_{DD}}{3(18+\sigma_{\nu^{\prime}})}
\left[(17+\sigma_{\nu^{\prime}}) +
\frac{(21+\sigma_{\nu^{\prime}})}{3} \frac{L}{B}+
\frac{2}{3}\frac{(9+5\sigma_{\nu^{\prime}})}{\sigma_{\nu^{\prime}}} \frac{L^{\prime}}{B}
\right] \ .
\end{eqnarray}
The results of the calculation are summarized in
Fig.~\ref{fig:Omega}.
This figure shows what are the allowed values of
the parameter $\xi$ defined below, as a function of the mass of
the LTB, for a given $T^*$, if the LTB accounts for the whole dark
matter density of the universe. The parameter $\xi$ can be
considered roughly speaking as the total ratio of lepton over
baryon number, with the new lepton family number $L'$ weighted
``appropriately'' due to the large mass that $\nu'$ and $\zeta$
carry.

\beq \xi \equiv \frac{L}{B} +
\frac{2}{\sigma_{\nu'}}\frac{9+5\sigma_{\nu'}}{21+\sigma_{\nu'}}\frac{L'}{B}
\ . \label{soxidef} \eeq

\begin{figure}[tp]
\begin{center}
\includegraphics[width=0.7\linewidth]{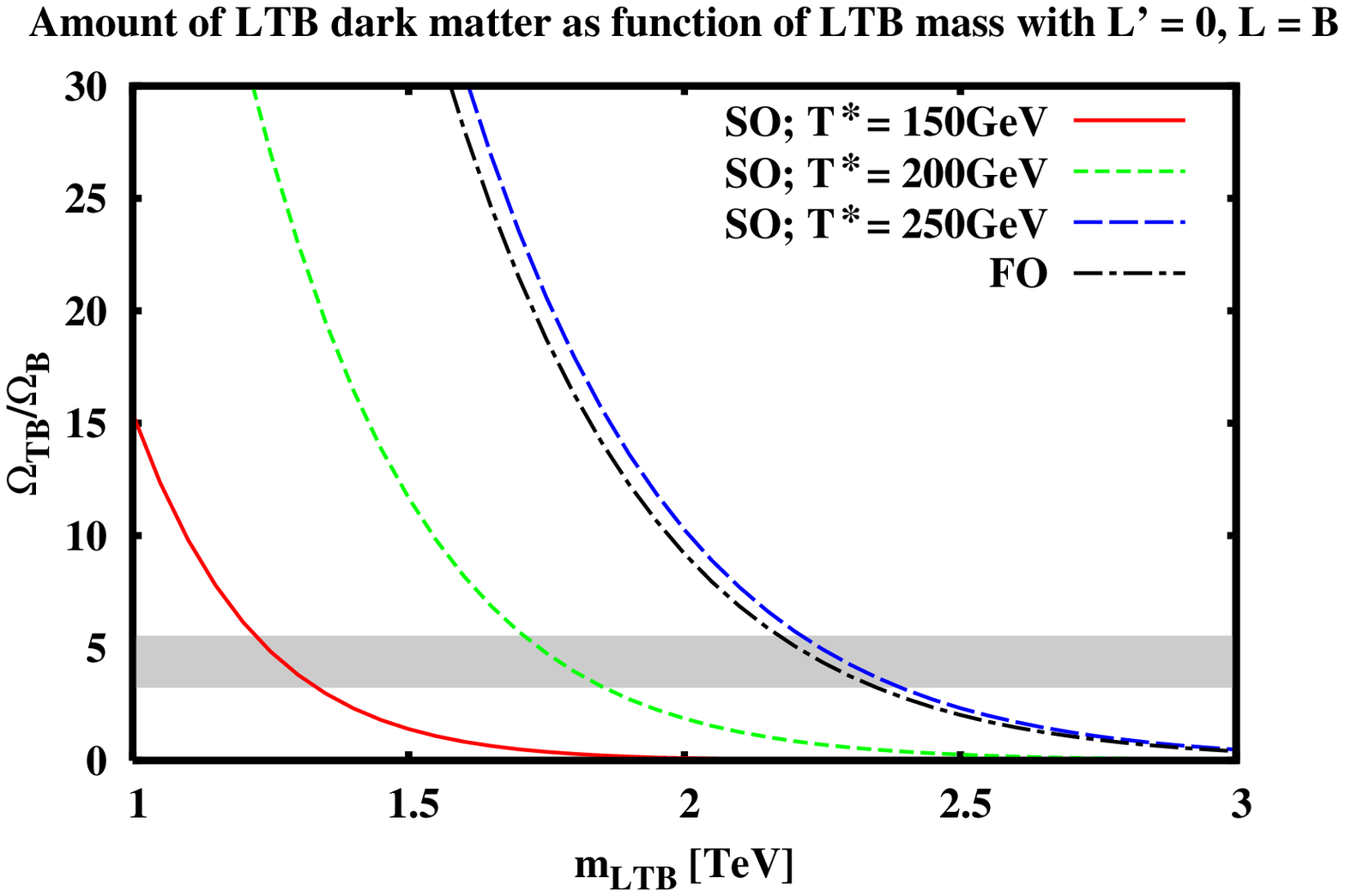}
\caption{Amount of LTB DM as function of the mass of the LTB
  particle. The plot is shown for $L'=0$ and $L=B$ for second order
  (SO) phase transitions with various temperatures $T^*$ and
  a for first
  order (FO) phase transition as well. }
\label{fig:amount}
\end{center}
\end{figure}

From Fig.~\ref{fig:Omega} we see for example that if we set $L'=0$
(no new leptons present) while also setting $L/B=-4$, we
need a mass for the LTB somewhere between 1.1 to 2.2
  TeV,
according to what is the freeze out temperature $T^*$. We should
emphasize that there are two branches of allowed values for $\xi$,
separated by the dotted horizontal line. The lower branch, as for
example the one we just described with $\xi=-4$, corresponds to
a relic density made by technibaryons DD. The upper set of allowed
values, (as for
$\xi=2$), corresponds to the DD antiparticle.
In Fig. \ref{fig:amount} we show the dependence of the neutral
technibaryon matter density as a function of its mass for a fixed
value of the parameter $\xi$. We see that if the LTB mass is lighter
than roughly a TeV the density of
the particle is very large,
giving
a too large ratio $\Omega_{TB}/\Omega_{B}$. So, for a given
value of $\xi$ and $T^*$, WMAP data put constraints on the allowed
mass of the technibaryon. On the other hand if we increase enough
the mass of the technibaryon, we can get a ratio less than 4-5,
which means that the technibaryon can be a component of the dark
matter density.

\subsection{1st Order Phase Transition}
If the electroweak phase transition is predicted to be first order,
then the baryon, lepton and technibaryon
violating processes ``freeze'' slightly above the phase
transition. For this reason, we have to impose two
conditions, the overall charge neutrality $Q=0$ and $Q_3=0$, where
$Q_3$ is the charge associated with the $T_3$
isospin generator of the weak interactions. This charge has to be zero
because above the phase transition the
electroweak symmetry is not broken and therefore $Q_3=0$.

\begin{figure}[tp]
\begin{center}
\includegraphics[width=0.7\linewidth]{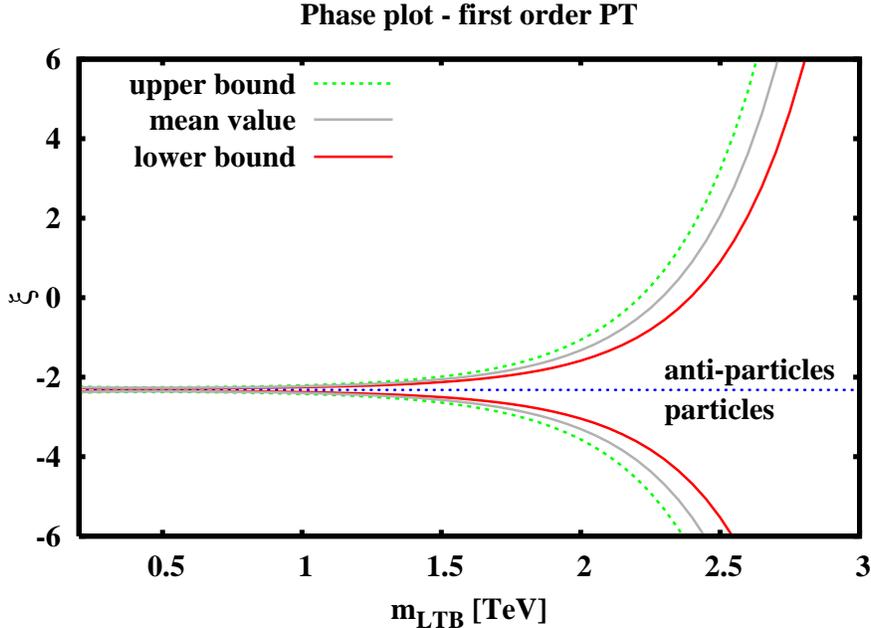}
\caption{Plot representing the region of the parameters according to
    which the fraction of technibaryon matter
    density over the baryonic one takes on the values
    $[3.23,\,5.55]$. Here we consider the case of a first order phase
    transition.
    The parameters in the plot are the mass of the LTB DM particle and
    $\xi$ of Eq.~(\ref{foxidef}).
    The dotted line separates areas of abundant particles
    and anti-particles. }
\label{fig:dmfoPT}
\end{center}
\end{figure}

The technibaryon over baryon number density ratio is, in the same
approximation used for the second order phase transition:
\begin{eqnarray}
-\frac{TB}{B}={\sigma_{DD}} \frac{22+\sigma_{\nu^{\prime}}}
{9(22+2\sigma_{DD}+\sigma_{\nu^{\prime}})}
\left[3+\frac{L}{B} +
  \frac{1}{\sigma_{\nu^{\prime}}}\frac{L^{\prime}}{B}\right] \ .
\end{eqnarray}
$T^*$ is expected to be larger than that of the second order case,
i.e. it should be identified with the critical temperature $T^c$
of the electroweak phase transition. This fact forces the mass of
the LTB to be larger than that of the second order case to describe the
whole DM. Our results are summarized in Figure \ref{fig:dmfoPT}.
As in the case of the 2nd order phase transition, we have plotted
the allowed values of the $\xi$ parameter
\beq \xi \equiv \frac{L}{B} +
  \frac{1}{\sigma_{\nu'}}\frac{L'}{B}
\ , \label{foxidef} \eeq
which is slightly
different from the previous case, as a function of the LTB mass,
under the WMAP constraints regarding the overall density of dark
matter in the universe. Using the previous example of $L'=0$ and
$L/B=-4$, we get an LTB mass of around 2.2 TeV.

\section{Detection of The Neutral Technibaryon}

Apart from the possibility of detecting a technibaryon in the LHC experiment it would be certainly interesting
to detect the neutral technibaryon in earth based experiments such as the CDMS \cite{Akerib:2004fq,Mandic:2004cr,Wang:2005cs,Akerib:2005kh}. Two are the basic ingredients affecting the detection of a cold DM object in these kinds of experiments. The first one is how large is
 the cross section of the object to be observed with the matter in the detector.
 The second has to do with the local density of
 DM in general and of the specific component of DM in particular. Current estimates suggest that the local
 density for a single component should be somewhere between
0.2-0.4 GeV/cm$^3$. It is evident that the higher the cross
section and the local density of dark matter are, the larger are
the chances for the detection of the particle. The CDMS
collaboration for example, can identify WIMPs by observing the
recoil energy produced in elastic scattering between the WIMP and
a nucleus in the detector. The expected rates of events per unit
time and mass of the detector has been calculated in several
places and we refer to the review paper \cite{Lewin:1995rx} for a
complete list of relevant references. The number of counts
reported by the detector per unit time, mass of the detector and
recoil energy is
 \beq
  \label{rate1}
\frac{dR}{dT}=\frac{R_0}{E_0 r}e^{-T/E_0r} \ ,
 \eeq
 where $T$ is
the recoil energy of the nucleus, $E_0$ is the kinetic energy of the WIMP and $r=4mM_n/(m+M_n)^2$, $m$ and $M_n$
being the masses of the WIMP and the nucleus respectively. The parameter $R_0$ is the total rate containing the
information about the cross section and is given by
  \beq
   R_0 =
\frac{2}{\pi^{1/2}}\frac{N_0}{A}\frac{\rho_{dm}}{m}\sigma_0
\upsilon_0 \ , \label{rate2} \eeq where  $N_0$ is the Avogadro
number, $A$ is the atomic number of the nucleus of the detector,
$\rho_{dm}$ is the local dark matter density, $\sigma_0$ is the
cross section for an elastic collision between the WIMP and the
nucleus and $\upsilon_0$ is the thermal velocity of the WIMPs. One
should note here that Eq.~(\ref{rate1}) is an approximate
expression. In reality the calculation is more elaborate. {}For
example, in principle one has to assume a Maxwell distribution for
the velocities of the WIMPs up to the escape velocity for our
galaxy. In addition, the effect of the motion of the earth
relatively to the halo should be considered. These factors can
change the expected rate. The total rate of counts can be more
usefully rewritten in convenient units as \beq R_0=\frac{503}{M_n
m}\left(\frac{\sigma_0}{1{\rm
 pb}}\right)\left(\frac{\rho_{dm}}{0.4{\rm GeV}{\rm
 cm}^{-3}}\right)\left(\frac{\upsilon_0}{230{\rm
    kms}^{-1}}\right)
\frac{{\rm GeV}^2}{{\rm kg}.{\rm days}} \ .\eeq

 Since our
prospective dark matter component is a Goldstone boson, we are
interested only in the spin independent elastic cross section.
This is given in natural units by \cite{Goodman:1984dc} \beq
\sigma_0=\frac{G_F^2}{2\pi}\mu^2 \bar Y^2 \bar{N}^2 F^2 \ , \eeq
where $G_F$ is the Fermi constant and ${\bar Y = 2Y}$. {}For a
Dirac fermion ${\bar Y = Y_L+Y_R}$ and $\mu$ is the reduced mass
of the latter and the nucleus target. $\bar{N} = N -
(1-4\sin^2\theta_w)Z$, where $N$ and $Z$ are the number of
neutrons and protons in the target nucleus and $\theta_w$ is the
Weinberg angle. The parameter $F^2$ is a form factor for the
target nucleus. The cross section can be written as
 \beq \sigma_0 = 8.431\times10^{-3}\frac{\mu^2}{{\rm GeV}^2} \bar Y^2 \bar{N}^2 F^2
\,{\rm
  pb} \ .
\label{rate3}\eeq The Ge atom has 41 neutrons and 32 protons,
giving an $\bar N = 38.59$. Our LTB has $\bar Y=1$ \footnote{We
have directly computed this value for $\bar Y$ using the effective
Lagrangian in \cite{Gudnason:2006ug}. {}For a generic $y$ we have
$\bar Y= 2y -1$ and it coincides with the value one deduces by
constructing the $LTB$ wave function as follows: $
\sqrt{2}\,|\,LTB\rangle = |\, D_LD_L\rangle - |\,D_RD_R \rangle
$.}. Since Standard Model neutrinos have $\bar Y=1/2$, this means
that the cross section for the technibaryon will be four times
larger than the one corresponding to a heavy neutrino. As we
already mentioned, for typical values of the $L/B$ ratio, in order
to get the whole density for the dark matter, the mass of the
technibaryon should be of the order of a TeV. The form factor
$F^2$ for the nucleus of Ge depends on the recoil energy $T$. It
models the loss of coherence of the scattering for large recoil
energies.
 For
typical values of the recoil energy around 20-50 keV one expects $F^2$ to be around $0.58$. We
estimated the nuclear form factor $F$ using the solid
sphere approximation -proper for the spin-independent WIMP
interaction- summarized in \cite{Lewin:1995rx}. To be more precise the
nuclear form factor ranges from $0.72$ to  $0.43$ when
the recoil energy ranges from 20 to 50 keV\footnote{ We may have overestimated the nuclear form factor \cite{Mandic:2004cr}. If the physical value of $F^2$ is lower than the one we used then the allowed fraction of LTB-DM increases.}.

The number of counts that are
detectable is given by \beq  {\rm counts} = \frac{dR}{dT}\Delta T \times \tau \ , \label{rate4} \eeq where
$\tau$ is the exposure of the detector measured in kg.days and $\Delta T$ is the energy resolution of the
detector. In the CDMS experiment a 19.4 kg.days exposure was achieved for the Ge detectors with an energy
resolution of $\Delta T = 1.5$~keV. So far no counts have been found. The $90\%$ level of confidence would lead
to 2.3 counts.

If we assume that our LTB constitutes the whole DM in the Universe we have seen from our previous computations
that a typical value of the mass is about 2 TeV, for the second order phase transition case. Taking a recoil
energy around 50 keV, $\rho_{dm}=0.3$ GeV/cm$^3$ and $F^2 \sim 0.43$ the number of counts predicted is
around 13 which is a value few times larger than the 90\% confidence value presented before. By stretching the
parameters we can reduce, or even annihilate the gap, between our
prediction and experimental bounds. Using
still a mass around 2 TeV but choosing a different set of input,
i.e. $\rho_{dm}=0.1$ GeV/cm$^3$, $F^2 \sim
0.3$ and $T=70$~{keV} one finds
around two predicted counts. Hence we
would be within the 90\% confidence
level. Under these rather extreme conditions one cannot yet completely exclude the possibility that our WIMP can
constitute the whole DM. Another simple way to reduce
the gap between experiment and our LTB particle,
if we imagine it to be a component of DM, is to increase its mass. In
doing so, however, we
neglect the relevant information gained in the previous sections in
which we related the mass of the LTB to the
fraction of DM in the Universe it can account for.

\begin{figure}[!tbp]
\begin{flushleft}
\hspace{6pt}\includegraphics[width=0.94\linewidth]{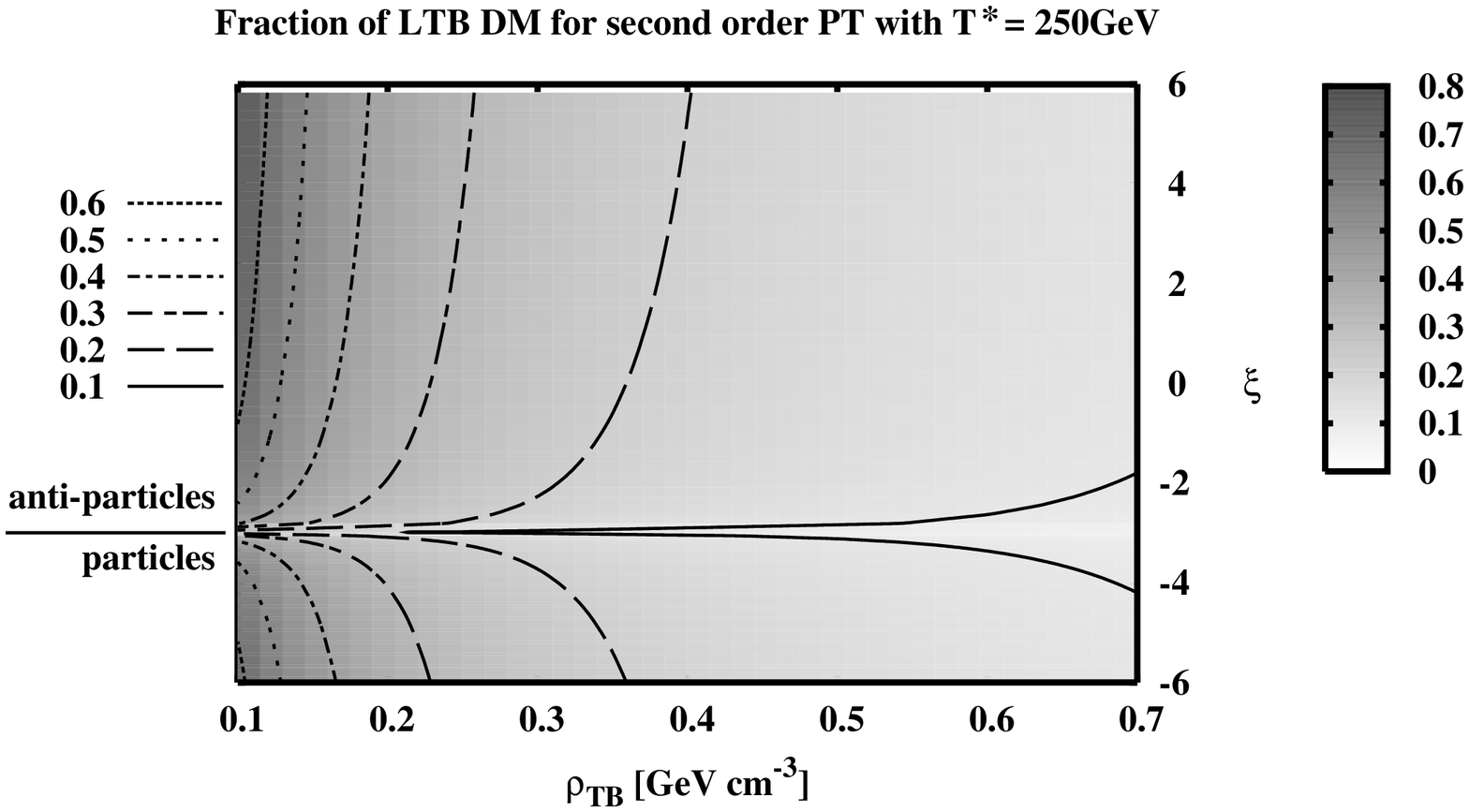}\\[0.5cm]
\includegraphics[width=\linewidth]{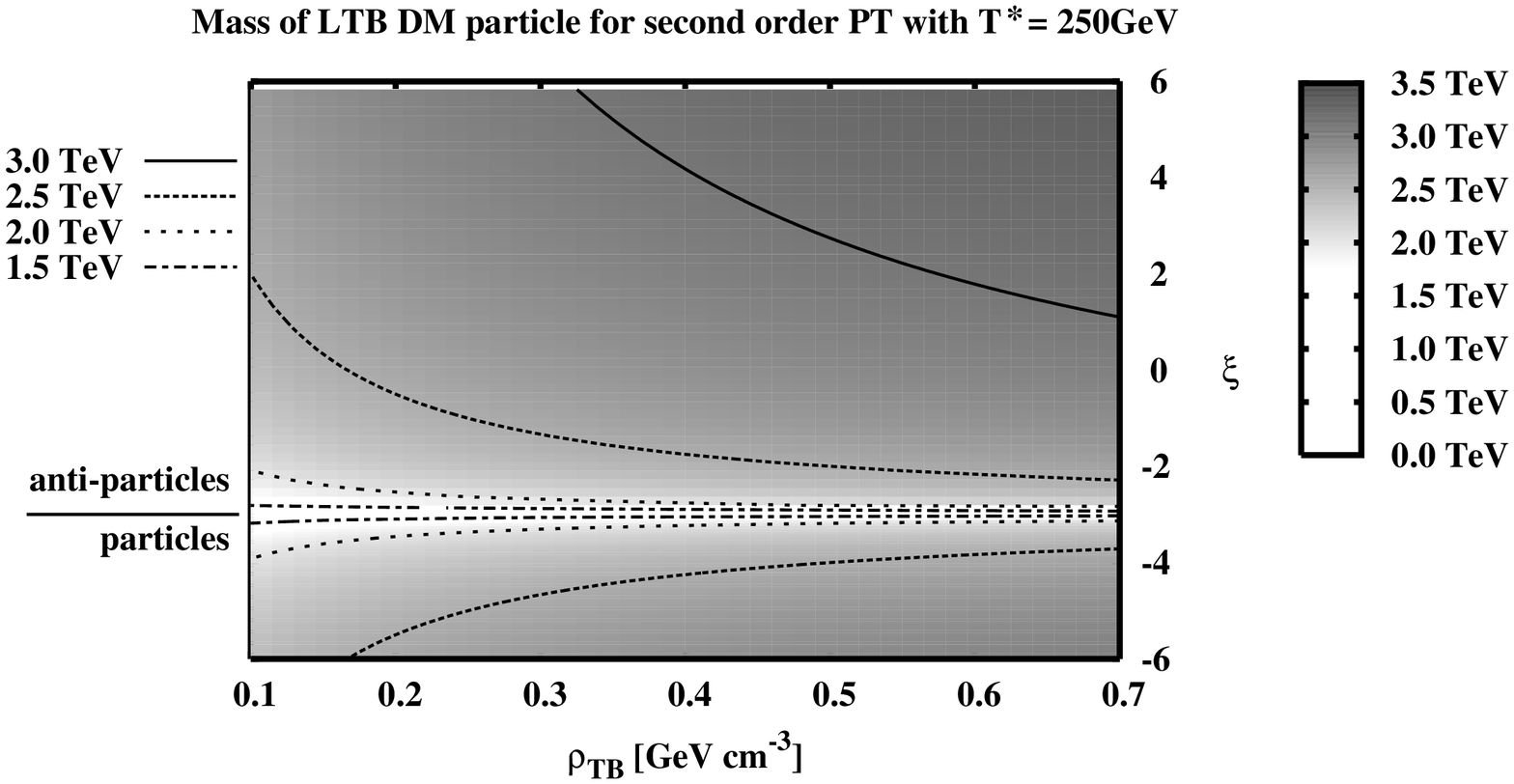}
\caption{\emph{Top Panel}: The maximal fraction of local DM
    density allowed by the 90\% experimental constraint as function of
    the local DM density and the parameter $\xi$ of
    Eq.~(\ref{soxidef}).
   \emph{Bottom Panel}: For the corresponding maximal fraction of local DM
   density currently allowed by the 90\% experimental constraint as
    function of the local DM density and $\xi$ we plot the associated
    LTB mass.
    Both plots are presented with second order phase
    transition with $T^* = 250$ GeV and a recoil energy $T=50$~keV.}
\label{universal1}
\end{flushleft}
\end{figure}

We now take into account, in a more careful way, such a dependence on the mass of the LTB. {}From the previous
section we learned that the general trend is that the amount of DM saturated by our LTB object decreases when
increasing the mass of the LTB. In the absence of a complete computation of how DM distributes itself in the
Universe we make the oversimplifying assumption that the fraction of local DM density of our LTB follows the
same fraction of DM in the Universe. At this point we impose the 90\% experimental constraint. Our results are
reported in Fig.~{\ref{universal1}}.
In the figure we have chosen $F^2 \simeq 0.43$ and the thermal
velocity is $230$~km/s. We present both the maximal fraction of
local DM density determined imposing the 90\% experimental
constraint (Fig.~\ref{universal1}) and the associated value of the
LTB mass as functions of $\xi$. We have allowed for variations
 of the parameters to make our analysis more complete. Note that we
  have allowed the local DM density to reach, in the plots, very large
   values although a more modest range (i.e. up to 0.4 {GeV}/cm$^3$) is probably sufficient.

Summarizing we can say that for reasonable values of the input
parameters the 90\% experimental constraint
allows for a 10\% to 65\% of LTB-DM component in the Universe. Our DM
component allowed mass ranges
between 1.4 and 3.3 TeV depending on the order of the associated electroweak phase transition as well as the exact
value of the local DM density and experimental parameters range. Our conclusions are slightly affected if we use 20 keV as recoil energy.
In this case at most we can account for 30\% of the DM in the Universe and
the masses of the LTB would be slightly higher.

  The question to be answered at this point is: What makes the rest of
  the DM in
  the Universe? We speculate that a
  techni-axion, needed for the
  solution of the strong CP problem, could be a natural candidate (see
  for example \cite{Davoudiasl:2006bt}). In
  this way the two components for DM are associated to two natural and
  complementary extensions of the SM.
  An explicit model containing axions from
  technicolor-like dynamics has been
  constructed in \cite{Hsu:2004mf}.

\section{Conclusions}

We have investigated in much detail dark matter candidates emerging in our recently proposed technicolor
theories. We have determined the contribution of the lightest, neutral, stable technibaryon to the dark matter
density having imposed weak thermal equilibrium conditions and overall electric neutrality of the Universe.
Sphaleron processes have been taken into account. We performed the analysis in the case of a first order
electroweak phase transition as well as a second order one. In both cases, the new technibaryon contributes to
the dark matter in the Universe. We have also examined the problem of the constraints from earth based
experiments. {We find that quite a substantial amount of DM can be explained within our model}. The new generation of DM-detection experiments can put very strict limits or even rule out the
present type of DM component. {We should stress that our framework can be applied to any model featuring a new baryonic type particle at the electroweak scale whose new baryon-type charge is violated only by weak interactions.}

\acknowledgments
It is a pleasure to thank D.D. Dietrich, D.K.
Hong, K. Tuominen and J.D. Vergados for discussions and careful
reading of the manuscript. The work of C.K. and F.S. is supported by the Marie Curie Excellence
Grant under contract MEXT-CT-2004-013510.
F.S. is also supported as Skou fellow of the Danish Research
Agency.

\end{document}